\documentclass[smallextended]{svjour3}
\usepackage{amssymb,amsmath}
\usepackage{graphicx}

\newcommand*{\kB}{k_{\mathrm{B}}} \newcommand*{\mC}{m_{\mathrm{C}}}
\newcommand*{\ust}{u_\ast} \newcommand*{\Ust}{U_\ast}
\newcommand*{\ubt}{u_\bot}
\newcommand*{\nd}{n_\#} \newcommand*{\Md}{M_\#}
\newcommand*{\mud}{\mu_\#}
\newcommand*{\rA}{\mathrm{A}} \newcommand*{\rN}{\mathrm{N}}
\newcommand*{\rI}{\mathrm{I}} \newcommand*{\rK}{\mathrm{K}}
\newcommand*{\bv}{\mathbf{v}} \newcommand*{\bV}{\mathbf{V}}
\newcommand*{\bw}{\mathbf{w}}
\newcommand*{\cL}{\mathcal{L}} \newcommand*{\cM}{\mathcal{M}}
\newcommand*{\cUcm}{\mathcal{U}_{\mathrm{cm}}}
\DeclareMathOperator{\erf}{erf}

\journalname{Rendiconti Lincei.\ Scienze Fisiche e Naturali}

\begin{document}

\title{Scattering of particles from a solid surface: The impulsive model of composite encounters}

\titlerunning{The impulsive model of composite encounters}

\author{Vyacheslav M.~Akimov \and Vladimir M.~Azriel \and Lyubov I.~Kolesnikova \and Lev Yu.~Rusin \and Mikhail B.~Sevryuk}

\institute{Vyacheslav M.~Akimov \email{vyacheslav-akimov@rambler.ru}
\and Vladimir M.~Azriel \email{azriel\_vladimir@mail.ru}
\and Lyubov I.~Kolesnikova \email{ek7787@rambler.ru}
\and Lev Yu.~Rusin \email{rusin@chph.ras.ru}
\and Mikhail B.~Sevryuk \email{2421584@mail.ru}
\at V.L.~Talrose Institute for Energy Problems of Chemical Physics, N.N.~Sem\"enov Federal Research Center of Chemical Physics, Russian Academy of Sciences, 38 Leninsky Prospect, Bld.~2, Moscow 119334, Russia}

\date{Received: date / Accepted: date}

\maketitle

\begin{abstract}
We propose a general impulsive model for scattering of molecules from a flat solid surface. It is assumed within the framework of this model that an encounter of an atom (or ion) with the surface is a series of elastic (in the direction normal to the surface) hits of the atom against surface pseudoparticles, the hits instantly following each other. To each atom, one assigns two infinite sequences of masses of pseudoparticles. The model is a far-reaching generalization of the well-known hard cube model. Criteria for both finiteness and infinity of series of hits are formulated, based on the masses of pseudoparticles and the mass of the atom. It is shown that in virtually all the cases, any number of hits in a series occurs with a positive probability. The proposed model does not satisfy the reciprocity condition.
\keywords{Scattering of atoms from a surface \and Impulsive model \and Surface pseudoparticles \and Series of hits}
\end{abstract}

\section{Introduction}
\label{introduction}

In the development of methods for theoretical description of scattering of atomic particles (atoms or ions) from a solid surface, an important milestone was the \emph{hard cube model} due to Logan and Stickney \cite{Logan66}, which has much in common with Goodman's earlier model \cite{Goodman65}. The Logan--Stickney model is \emph{impulsive}: the atom--surface interaction in this model is considered as an elastic hit of the atom against a certain pseudoparticle (``cube'') which is moving normally to the surface assumed to be flat, while the component of the atom velocity parallel to the surface is conserved. Impulsive models of scattering of atoms or molecules from a surface, developing the hard cube model, have been used in chemical physics until very recently (see, e.g., \cite{Azriel18,Azriel19,Dorenkamp18,Liang18,Majumder18,Sipkens17}). The role of simple models and, in particular, of Logan--Stickney-type models in the contemporary theory of gas--surface interactions is discussed in the survey \cite{Kleyn08}. An impulsive treatment of scattering of molecules at the gas--liquid interface is exemplified by the recent paper \cite{Livingston19}.

Several key generalizations of the hard cube model are as follows. In the article \cite{Grimmelmann80}, this model is augmented by an attractive step-like potential. In the \emph{soft cube model} \cite{Logan68}, the surface atoms are represented by one-dimensional harmonic oscillators which vibrate normally to the surface, and the interaction of a gas atom with an oscillator is described by a potential that consists of an exponential repulsive part and a subsequent finite step. In the \emph{hard spheroid model} \cite{Steinbruchel80,Steinbruchel82}, the surface pseudoparticles are provided with spherical ``caps'', and the gas atoms interact with these ``caps'' in an impulsive manner. In the very popular \emph{washboard model} \cite{Tully90} (see also \cite{Liang18}) close to the hard spheroid model, one also postulates an impulsive interaction of gas atoms with surface pseudoparticles, but the surface itself is sinusoidally corrugated. The component of the atom velocity normal to the surface at the point of impact undergoes an instant change, while the velocity component parallel to the tangent to the surface at that point is conserved. The model also includes an attractive step-like potential. Corrugated surfaces with folds of other shapes are considered in the paper \cite{Xia94}. The article \cite{Yan04} proposed a generalization of the washboard model where the surface pseudoparticles are characterized not only by mass but also by moment of inertia (but the component of the atom velocity parallel to the tangent to the surface at the point of impact is still conserved). The paper \cite{Mateljevic09} is devoted to another generalization of the washboard model where the surface is described by a finite collection of circular Gaussian ``hills'' and ``valleys'' with random location and random height/depth. In the model of the paper \cite{Doll73}, the gas above the surface consists of rigid rotors (representing homonuclear diatomic molecules) rather than of point masses, the rotations being constrained to be in planes orthogonal to the surface.

In this note, we propose one more generalization of the hard cube model, the so called \emph{composite encounter model}. In this model, an encounter of an atomic particle with a flat surface is a series of hits of the atom against surface pseudoparticles, the hits instantly following each other and being elastic in the direction normal to the surface. In our opinion, the composite encounter model is sufficiently flexible (the law of the change in the atom velocity component normal to the surface includes infinitely many parameters) and possesses a rather interesting mathematical structure. A preliminary version of the model was expounded in Russian in the work \cite{Azriel19}, while its particular case pertaining to ionic dissociation of potassium iodide molecules at a graphite surface was presented in the paper \cite{Azriel18}. A simulation of this dissociation via the composite encounter model within the framework of the quasiclassical trajectory method \cite{Azriel18} and the additional calculations of the work \cite{Azriel19} have confirmed that the model is very easy to implement for computers.

The paper is organized as follows. We briefly describe the hard cube model in Sect.~\ref{HardCube} and explicate the composite encounter model in Sect.~\ref{CompositeEncounter}. In Sect.~\ref{numberofhits} we show that in almost all situations, the number of hits of the atom against surface pseudoparticles in a series can be any positive integer with a nonzero probability. Criteria for finiteness of a series of hits are formulated in Sect.~\ref{finite} and those for infinity are given in Sect.~\ref{infinite}. In Sect.~\ref{reciprocity} we show that the reciprocity condition is violated for the composite encounter model. Finally, simulation of scattering of systems of atoms (of molecules) from a surface via combining the quasiclassical trajectory method and the composite encounter model is discussed in Sect.~\ref{conclusion}.

The results of Sects.~\ref{finite}--\ref{reciprocity} were partly published in Russian in the paper \cite{Azriel19}.

\section{Hard cube model}
\label{HardCube}

Within the framework of the hard cube model \cite{Logan66}, the surface is assumed to be the infinite ``horizontal'' plane $z=0$ (where $x,y,z$ are the Cartesian coordinates in the space) and has temperature $T_s$ (measured in Kelvin). The atom (or ion) of mass $m$ moves in the half-space $z>0$, so that a negative value of the $z$-component of the atom velocity (this component is orthogonal to the surface) corresponds to approaching the surface while a positive value corresponds to moving away from the surface. As was already noted in the introduction, the interaction of the atom with the surface is treated as an elastic hit against a certain \emph{pseudoparticle} of mass $M$, its velocity being ``vertical'' (i.e., normal to the surface). To be more precise, suppose that just before an encounter of the atom with the surface, the $z$-component of the atom velocity is equal to $u<0$. As a result of the encounter, the component of the atom velocity parallel to the surface (i.e., the sum of the $x$- and $y$-components) remains unaltered while the $z$-component changes instantly. The new value $u'$ of the ``vertical'' component of the atom velocity within the framework of the hard cube model is determined by the following algorithm.

1) One chooses the absolute value $\Ust\geq 0$ of the velocity of the surface pseudoparticle just before the encounter at random from the one-dimensional Maxwellian distribution corresponding to the temperature $T_s$ and mass $M$. Otherwise speaking, the probability density function of the random variable $\Ust$ is
\[
F_{M/T_s}(\Ust) = \left(\frac{2M}{\pi\kB T_s}\right)^{1/2}\exp\left(-\frac{M\Ust^2}{2\kB T_s}\right),
\]
where $\kB$ is the Boltzmann constant. The particle velocity itself is equal to $U=\pm\Ust$.

Note that the probability that $\Ust\leq\Gamma$ for some positive number $\Gamma$ is equal to
\begin{equation}
\left(\frac{2M}{\pi\kB T_s}\right)^{1/2} \int_0^\Gamma \exp\left(-\frac{M\Ust^2}{2\kB T_s}\right) d\Ust = \erf\left(\left[\frac{M}{2\kB T_s}\right]^{1/2}\Gamma\right),
\label{erf}
\end{equation}
where
\[
\erf(t) = \frac{2}{\pi^{1/2}} \int_0^t e^{-\tau^2} d\tau
\]
is the error function which increases monotonously from $0$ to $1$ as $t$ grows from $0$ to $+\infty$.

2) The sign of the velocity $U$ (for $\Ust>0$) is chosen according to the following rule. Denote by $\ust=|u|=-u$ the absolute value of the $z$-component of the atom velocity just before the encounter. If $\Ust\geq\ust$ then the surface pseudoparticle is moving toward the atom ($U=\Ust$). On the other hand, if $0<\Ust<\ust$ then the pseudoparticle is moving toward the atom ($U=\Ust$) with probability $(\ust+\Ust)/(2\ust)>1/2$ and the pseudoparticle is moving in the same direction as the atom is ($U=-\Ust$) with probability $(\ust-\Ust)/(2\ust)<1/2$. In other words, $U>u=-\ust$ (the condition that a hit is possible) in all the cases, and the probability density function of the random variable $U$ is
\begin{equation}
\Phi_{M/T_s,\ust}(U) = \left[\begin{aligned}
F_{M/T_s}(U)&, & &\text{if $U\geq\ust$}, \\
\frac{\ust+U}{2\ust} F_{M/T_s}(U)&, & &\text{if $-\ust<U<\ust$}.
\end{aligned}\right.
\label{Phi}
\end{equation}
The function~\eqref{Phi} is continuous in the whole infinite interval $-\ust<U<+\infty$ and can be expressed by a single formula:
\[
\Phi_{M/T_s,\ust}(U) = \min\left(\frac{\ust+U}{2\ust}, \, 1\right) F_{M/T_s}(U).
\]

3) An instant elastic hit of the atom against the surface pseudoparticle occurs. As a result of this hit, the $z$-component of the atom velocity becomes equal to
\[
u' = 2\cUcm-u = \frac{(m-M)u+2MU}{m+M} > u,
\]
where $\cUcm = (mu+MU)/(m+M)$ is the ``vertical'' component of the velocity of the center of mass of the system ``the atom~-- the surface pseudoparticle''. As the velocity $U$ grows from $u$ to $+\infty$, the velocity $u'$ also increases from $u$ to $+\infty$. Note that
\[
u'-u = \frac{2M}{m+M}(U-u).
\]

For the atom to fly off the surface, the value of $u'$ should be positive. Logan and Stickney \cite{Logan66} assume that $M>3m$ (in the key section \mbox{IV-A} of their paper) or that $M>m$ (in Sect.\ \mbox{IV-B}) and consider mainly the averaged quantities
\[
\langle u'\rangle = \frac{(m-M)u+2M\langle U\rangle}{m+M}
\]
(in fact, averaging over $u$ is also carried out in their work, based on the temperature $T_g$ of the gas beam above the surface). This way they bypass an analysis of the situation where $u'\leq 0$ (this situation may be somewhat roughly interpreted as adsorption of the atom by the surface). At the same time, the probability of such a situation, i.e., the probability that the velocity $U$ lies in the interval $u<U\leq (M-m)u/(2M)$, is positive for any values of $m$, $M$, $T_s$, and $u$.

\section{Composite encounter model}
\label{CompositeEncounter}

The generalization of the hard cube model proposed in the present paper can be called the composite encounter model. The main idea of this model of scattering of atomic particles from a solid surface is that in the situation where $u'\leq 0$, one should complicate the procedure rather than ignore this inequality. The model is based on the following principles.

a) An encounter of an atom (or of an ion) with a surface is a \emph{series} of hits of the atom against surface pseudoparticles, the hits instantly following each other and the surface being given by the equation $z=0$. The change in the $z$-component of the atom velocity due to the $n$th hit corresponds to an elastic hit against a particle having mass $M_n$ and ``vertical'' velocity $U_n$ (the model is impulsive in the ``vertical'' direction). The series terminates as soon as the $z$-component of the atom velocity becomes positive (i.e., the atom flies off the surface).

b) The absolute value of the velocity $U_n$ has the one-dimensional Maxwellian distribution corresponding to surface temperature $T_s$ and mass $\mu_n$, the latter is not assumed to coincide with $M_n$.

c) The masses $M_n$ for different numbers $n$ are not assumed to coincide either, and the same is valid for the masses $\mu_n$ for different numbers $n$. The velocities $U_n$ for different numbers $n$ are chosen independently: the velocity $U_n$ for $n\geq 2$ is not connected in any way with the ``vertical'' velocity that the surface pseudoparticle acquires as a result of the $(n-1)$th hit.

d) The ``horizontal'' (i.e., parallel to the surface) component of the atom velocity can change upon any hit (the model allows tangential forces acting on the atom).

Now let us give a formal description of the composite encounter model. As in the hard cube model, the surface within the framework of the composite encounter model is the infinite ``horizontal'' plane $z=0$ and has temperature $T_s$, while the atom of mass $m$ moves in the half-space $z>0$ (the values of $m$ and $T_s$ are positive and finite). The interaction of the atom with the surface is characterized by two infinite sequences of masses, namely, by ``collisional'' masses $M_1,M_2,M_3,\ldots$\ and by ``Maxwellian'' masses $\mu_1,\mu_2,\mu_3,\ldots$\ as well as by an infinite sequence $\cL_1,\cL_2,\cL_3,\ldots$\ of laws of the change in the ``horizontal'' component of the atom velocity. Each of the masses $M_n$ and $\mu_n$ ($n\geq 1$) is greater than zero but may be infinite: $0<M_n\leq+\infty$, $0<\mu_n\leq+\infty$.

An encounter of the atom with the surface is a series of hits of the atom against surface pseudoparticles, the hits instantly following each other. Before the $n$th hit, the atom has a certain ``vertical'' velocity $u_{n-1}\leq 0$ (moreover, $u_0<0$) and a certain ``horizontal'' velocity $\bv_{n-1}$. The ``vertical'' velocity $U_n$ of the surface pseudoparticle before the $n$th hit is determined as follows. If $\mu_n=+\infty$ then $U_n=0$. If $\mu_n$ is finite and $u_{n-1}=0$, then $U_n\geq 0$ and $U_n$ is chosen at random from the one-dimensional Maxwellian distribution corresponding to the temperature $T_s$ and mass $\mu_n$. Otherwise speaking, the probability density function of the random variable $U_n$ is
\[
F_{\mu_n/T_s}(U_n) = \left(\frac{2\mu_n}{\pi\kB T_s}\right)^{1/2}\exp\left(-\frac{\mu_nU_n^2}{2\kB T_s}\right).
\]
Finally, if $\mu_n$ is finite and $u_{n-1}<0$, then $U_n>u_{n-1}$ and $U_n$ is chosen in the same way as in the hard cube model but with $M$ replaced by $\mu_n$. Otherwise speaking, the probability density function of the random variable $U_n$ is
\[
\Phi_{\mu_n/T_s,|u_{n-1}|}(U_n) = \min\left(\frac{|u_{n-1}|+U_n}{2|u_{n-1}|}, \, 1\right) F_{\mu_n/T_s}(U_n).
\]
The probability density function of the variable $|U_n|$ here is $F_{\mu_n/T_s}(|U_n|)$. Note that $U_n>u_{n-1}$ (the condition that a hit is possible) in all the cases, except for the situation where $U_n=u_{n-1}=0$.

As a result of the $n$th hit, the ``vertical'' component of the atom velocity instantly becomes equal to
\begin{equation}
u_n = \frac{(m-M_n)u_{n-1}+2M_nU_n}{m+M_n}.
\label{hit}
\end{equation}
The formula~\eqref{hit} is written both for a finite mass $M_n$ (if this is the case, this formula should be taken literally) and for the case where $M_n=+\infty$ (and this formula turns into $u_n = 2U_n-u_{n-1}$). In the sequel, we will not make such clarifications. Note that if $U_n>u_{n-1}$ then $u_n>u_{n-1}$, whereas if $U_n=u_{n-1}=0$ then $u_n=0$ as well. If $u_n>0$ then the series of hits terminates, the atom flies off the surface, and $u_n$ is the final value of the $z$-component of the atom velocity. On the other hand, if $u_n\leq 0$ then the $n$th hit of the atom against a surface pseudoparticle is instantly followed by the $(n+1)$th hit.

Note that if $\mu_n=+\infty$ then $u_n = (m-M_n)u_{n-1}/(m+M_n)$ is uniquely determined by $u_{n-1}$. In the case where $\mu_n=+\infty$ for $1\leq n\leq\nu$, we will use the notation
\[
\psi_\nu = \prod_{n=1}^\nu \frac{m-M_n}{m+M_n}.
\]
In addition, we will always assume that $\psi_0=1$.

The ``horizontal'' component $\bv_n$ of the atom velocity after the $n$th hit is determined according to the law $\cL_n$ taking into account the ``horizontal'' component $\bv_{n-1}$ of the atom velocity before the hit and probably the velocities $u_{n-1}$ and $U_n$. The law $\cL_n$ may depend on certain parameters (for example, on the masses $m$, $M_n$, $\mu_n$ or on the temperature $T_s$) and may include generating sample values of certain random variables (for instance, of the ``horizontal'' velocity $\bV_n$ of the surface pseudoparticle). In this article, we do not specify the laws $\cL_1,\cL_2,\cL_3,\ldots$\ because we will be primarily interested in the criteria for finiteness or infinity of a series of hits of the atom against surface pseudoparticles, as well as in the number of hits in a series in the case where the series in question is finite. These aspects of an encounter of the atom with the surface are connected exclusively with the changes in the $z$-component of the atom velocity and bear no relation to the ``horizontal'' component. We only note that if $\bV_n$ is collinear to $\bv_{n-1}$ (in particular, if $\bV_n=0$), then it is natural to assume that $\bv_n$ is also collinear to $\bv_{n-1}$. The simplest law $\cL_n$ is that the ``horizontal'' component of the atom velocity is not altered: $\bv_n=\bv_{n-1}$.

In the case where the series of hits is finite, we will denote by $\ubt>0$ the final value of the $z$-component of the atom velocity. If the number of hits in the series is equal to $\ell$ then $\ubt=u_\ell$.

Note that if one leaves the ``horizontal'' components of the atom velocities ``behind the scenes'', then a simultaneous $d$ times increase in all the ``collisional'' masses $M_1,M_2,M_3,\ldots$\ ($0<d<+\infty$) is equivalent to a $d$ times decrease in the mass $m$, and a simultaneous $d$ times increase in all the ``Maxwellian'' masses $\mu_1,\mu_2,\mu_3,\ldots$\ is tantamount to a $d$ times decrease in the temperature $T_s$.

It is interesting that in the early paper \cite{Goodman65}, a collision of an atom with a surface was also described as a series of elastic hits (which follow each other) of the gas atom against surface pseudoparticles (atoms). It is assumed in the article \cite{Goodman65} that both the gas atoms and the surface atoms move normally to the surface, the surface atom being contained in a one-dimensional ``box'' impermeable to it (but absolutely permeable to the gas atom) and bouncing elastically from the ends of the ``box''. The impacts ``from above'' of the gas atom on the surface atom are also elastic. Unlike in our model, the velocity $U_n$ of the surface atom before the $n$th hit ($n\geq 2$) of the two atoms is uniquely determined by the velocity $U'_{n-1}$ of the surface atom after the $(n-1)$th hit: $U_n=-U'_{n-1}$ (the surface atom has reflected from the ``lower'' end of the ``box''). The paper \cite{Yan04} examines in detail multiple encounters of an atom with a surface, but of a completely different nature. Each encounter is a single hit of the atom against a surface pseudoparticle (with a nonzero moment of inertia), but due to a pronounced corrugation of the surface, the atom after having flown off the surface may collide with it again at another point.

\section{Number of hits in a series}
\label{numberofhits}

In estimating the probabilities in this section, we will consider the masses $m$, $M_n$, $\mu_n$ ($n\geq 1$) as fixed (and satisfying certain conditions).

The possible number of hits of the atom against surface pseudoparticles in the composite encounter model is determined by the following trichotomy.

(a) Let $\mu_n=+\infty$ for all $n\geq 1$ and $\psi_n\geq 0$ for all $n$. Then $u_n=\psi_nu_0\leq 0$ for all $n\geq 1$. Thus, in this case the series of hits is infinite for any $T_s>0$ and $u_0<0$.

(b) Let $\mu_n=+\infty$ for $1\leq n\leq\nu$ ($\nu$ being some positive integer), $\psi_n>0$ for $1\leq n\leq\nu-1$, and $\psi_\nu<0$. Then $u_n=\psi_nu_0<0$ for $1\leq n\leq\nu-1$ and $u_\nu=\psi_\nu u_0>0$. Thus, in this case the series of hits includes exactly $\nu$ hits for any $T_s>0$ and $u_0<0$. Moreover, $\ubt=\psi_\nu u_0$ is uniquely determined by $u_0$.

(c) Let $\mu_n=+\infty$ for $1\leq n\leq\nu-1$ ($\nu$ being some positive integer) and $\psi_n\geq 0$ for $1\leq n\leq\nu-1$, and let the mass $\mu_\nu$ be finite. Then $u_n=\psi_nu_0\leq 0$ for $1\leq n\leq\nu-1$, so that for any $T_s>0$ and $u_0<0$, the series of hits includes at least $\nu$ hits. Moreover, in this case for any $T_s>0$ and $u_0<0$ and for any interval $[X,Y]$ with $0\leq X<Y<+\infty$, with a positive probability the number of hits in the series is exactly $\nu$ and $\ubt=u_\nu$ lies inside the interval $[X,Y]$.

Indeed, in this case $u_{\nu-1}=\psi_{\nu-1}u_0\leq 0$ is uniquely determined by $u_0$, and
\begin{equation}
u_\nu = \frac{(m-M_\nu)u_{\nu-1}+2M_\nu U_\nu}{m+M_\nu}.
\label{nu1}
\end{equation}
Consider arbitrary numbers $Y>X\geq 0$. If $X<u_\nu<Y$ then the number of hits is equal to $\nu$. The inequality $X<u_\nu<Y$ is equivalent to the fact that
\begin{equation}
\frac{(m+M_\nu)X+(M_\nu-m)u_{\nu-1}}{2M_\nu} < U_\nu < \frac{(m+M_\nu)Y+(M_\nu-m)u_{\nu-1}}{2M_\nu},
\label{XYnu}
\end{equation}
the left end of the range~\eqref{XYnu} being no smaller than $(M_\nu-m)u_{\nu-1}/(2M_\nu) \geq u_{\nu-1}$. For any $T_s>0$ and $u_{\nu-1}\leq 0$, the probability that $U_\nu$ falls in the interval~\eqref{XYnu} is positive for any finite mass $\mu_\nu$.

In particular, if the probability density function of the velocity $\ubt$ for given values of $T_s$ and $u_0$ is not a delta function, then it cannot identically vanish on any interval $[X,Y]$ with $0\leq X<Y<+\infty$.

If $\psi_{\nu-1}=0$ in the framework of case~(c) (this is only possible for $\nu\geq 2$), then the number of hits in the series is $\nu$ with probability $1$ for any temperature $T_s>0$. Indeed, in this situation $u_{\nu-1}=0$ and $u_\nu = 2M_\nu U_\nu/(m+M_\nu)$ is positive with probability $1$. On the other hand, if $\psi_{\nu-1}>0$ (so that the quantities $\psi_n$ for all $n$ less than $\nu-1$ are also other than zero, i.e., are positive, taking into account the context of case~(c)), then for a fixed temperature $T_s>0$ the probability that the number of hits in the series is $\nu$ tends to $1$ as $u_{\nu-1}\to 0$, i.e., as $u_0\to 0$. Indeed, in virtue of Eq.~\eqref{nu1}, the inequality $u_\nu>0$ is equivalent to the fact that
\begin{equation}
U_\nu > \frac{M_\nu-m}{2M_\nu}u_{\nu-1}.
\label{nu2}
\end{equation}
On the other hand, the inequality~\eqref{nu2} is definitely valid if
\begin{equation}
|U_\nu| > \max\left(\frac{|M_\nu-m|}{2M_\nu}, \, 1\right)|u_{\nu-1}|.
\label{nu3}
\end{equation}
For a fixed temperature $T_s$ and for the mass $\mu_\nu<+\infty$, the probability of the inequality~\eqref{nu3} holding tends to $1$ as $u_{\nu-1}\to 0$.

In particular, if the mass $\mu_1$ is finite then for any temperature $T_s>0$, the probability that the atom flies off the surface after the first hit against a pseudoparticle tends to $1$ as $u_0\to 0$.

Now suppose that in the framework of case~(c) $\psi_n>0$ for $1\leq n\leq\nu-1$ while the masses $\mu_n$ are finite for all $n$ in the range $\nu\leq n\leq N$ where $N\geq\nu$ is some positive integer. Then for any temperature $T_s>0$, any integer $\ell$ in the range $\nu\leq\ell\leq N$, any interval $[X,Y]$ with $0\leq X<Y<+\infty$, and any interval $[a,b]$ with $-\infty<a<b<0$ there exists a number $p$ in the range $0<p<1$ possessing the following property. The probability that the number of hits in the series is equal to $\ell$ and $\ubt=u_\ell$ lies inside the interval $[X,Y]$ is no less than $p$ for any velocity $u_0$ in the interval $[a,b]$.

Indeed, the inequality $a\leq u_0\leq b$ is tantamount to the inequality $a'\leq u_{\nu-1}\leq b'$, where $a'=\psi_{\nu-1}a$, $b'=\psi_{\nu-1}b$, and $-\infty<a'<b'<0$. For $n=\nu,\nu+1,\ldots,\ell-1$, require that the inequality
\[
\frac{2u_{n-1}}{3} \leq u_n = \frac{(m-M_n)u_{n-1}+2M_nU_n}{m+M_n} \leq \frac{u_{n-1}}{3}
\]
hold, i.e., require that the inequality
\begin{equation}
\frac{5M_n-m}{6M_n}u_{n-1} \leq U_n \leq \frac{2M_n-m}{3M_n}u_{n-1}
\label{M5623}
\end{equation}
be valid; note that the left end of the range~\eqref{M5623} is greater than $u_{n-1}$ for $u_{n-1}<0$. If
\[
\left(\frac{2}{3}\right)^{n-\nu}a' \leq u_{n-1} \leq \frac{1}{3^{n-\nu}}b',
\]
then the probability that $U_n$ falls in the interval~\eqref{M5623} for a fixed temperature $T_s>0$ and for the mass $\mu_n<+\infty$ is no less than a certain positive number $q_n<1$ (because this probability depends continuously on $u_{n-1}$ and is positive for any $u_{n-1}<0$). If $U_n$ lies in the range~\eqref{M5623} then
\[
\left(\frac{2}{3}\right)^{n-\nu+1}a' \leq u_n \leq \frac{1}{3^{n-\nu+1}}b'.
\]
Finally, given numbers $Y>X\geq 0$, require that the inequality
\[
X < u_\ell = \frac{(m-M_\ell)u_{\ell-1}+2M_\ell U_\ell}{m+M_\ell} < Y
\]
hold, i.e., require that the inequality
\begin{equation}
\frac{(m+M_\ell)X+(M_\ell-m)u_{\ell-1}}{2M_\ell} < U_\ell < \frac{(m+M_\ell)Y+(M_\ell-m)u_{\ell-1}}{2M_\ell}
\label{XYell}
\end{equation}
be valid (cf.\ Eq.~\eqref{XYnu}), then the number of hits in the series in question will be equal to $\ell$. Note that the left end of the range~\eqref{XYell} is no less than $(M_\ell-m)u_{\ell-1}/(2M_\ell) > u_{\ell-1}$. If
\[
\left(\frac{2}{3}\right)^{\ell-\nu}a' \leq u_{\ell-1} \leq \frac{1}{3^{\ell-\nu}}b',
\]
then the probability that $U_\ell$ falls in the interval~\eqref{XYell} for a fixed temperature $T_s>0$ and for the mass $\mu_\ell<+\infty$ is no less than a certain positive number $Q_\ell<1$. One can set $p$ to be the product $q_\nu q_{\nu+1}\cdots q_{\ell-1}Q_\ell$.

In particular, if all the ``Maxwellian'' masses $\mu_1,\mu_2,\mu_3,\ldots$\ are finite then for any temperature $T_s>0$ and velocity $u_0<0$ and for any positive integer $\ell$, a series of hits of the atom against surface pseudoparticles consists of exactly $\ell$ hits with a positive probability. Moreover, for any temperature $T_s>0$, any positive integer $\ell$, any interval $[X,Y]$ with $0\leq X<Y<+\infty$, and any interval $[a,b]$ with $-\infty<a<b<0$ there exists a number $p$ in the range $0<p<1$ possessing the following property. The probability that the number of hits in the series is equal to $\ell$ and $\ubt=u_\ell$ lies inside the interval $[X,Y]$ is no less than $p$ for any velocity $u_0$ in the interval $[a,b]$.

\section{Finite series of hits}
\label{finite}

As we saw in the previous section, if all the ``Maxwellian'' masses $\mu_1,\mu_2,\mu_3,\ldots$\ are finite then a series of hits of the atom against surface pseudoparticles can be arbitrarily long with a positive probability. At the same time, for the practical use of the composite encounter model, it is extremely important to know under what conditions on the masses $m$, $M_1,M_2,M_3,\ldots$\ and $\mu_1,\mu_2,\mu_3,\ldots$, for any temperature $T_s>0$ of the surface and for any initial ``vertical'' velocity $u_0<0$ of the atom the series of hits of the atom against surface pseudoparticles cannot be infinite. We will point out two situations where a series of hits is finite \emph{for sure} (i.e., with probability $1$).

1) The first situation consists in that the sequences $\{1/M_n\}_{n\geq 1}$ and $\{\mu_n\}_{n\geq 1}$ are bounded from above, i.e., there are positive constants $\Md$ and $\mud<+\infty$ such that $M_n\geq\Md$ and $\mu_n\leq\mud$ for all $n\geq 1$. Indeed, suppose that in this setup, all the velocities $u_n$ ($n\geq 0$) are non-positive and the series of hits is infinite. Fix an arbitrary number $\Gamma>0$. According to Eq.~\eqref{erf}, the probability that $|U_n|\geq\Gamma$ for a given number $n$ is equal to
\[
1-\erf\left(\left[\frac{\mu_n}{2\kB T_s}\right]^{1/2}\Gamma\right) \geq 1-\erf\left(\left[\frac{\mud}{2\kB T_s}\right]^{1/2}\Gamma\right) > 0.
\]
On the other hand, the probability that $U_n\geq\Gamma$ exceeds half of this quantity. Consequently, the sequence $U_1,U_2,U_3,\ldots$\ surely contains infinitely many terms not less than $\Gamma$, and therefore
\[
\sum_{n=1}^\infty \max(U_n,0) = +\infty.
\]
Since $M_n\geq\Md$, $u_{n-1}\leq 0$, and $U_n\geq u_{n-1}$ for all $n\geq 1$, we have
\begin{multline*}
u_n-u_0 = \sum_{\iota=1}^n (u_\iota-u_{\iota-1}) = \sum_{\iota=1}^n \frac{2M_\iota}{m+M_\iota}(U_\iota-u_{\iota-1}) \geq {} \\
\frac{2\Md}{m+\Md} \sum_{\iota=1}^n (U_\iota-u_{\iota-1}) \geq \frac{2\Md}{m+\Md} \sum_{\iota=1}^n \max(U_\iota,0),
\end{multline*}
and the last expression tends to infinity as $n\to\infty$. Consequently, $u_n>0$ for $n$ sufficiently large. The contradiction obtained shows that the probability of an infinite series of hits is $0$.

2) The second situation is that $M_n\geq m$ and $\mu_n<+\infty$ for all $n$ sufficiently large (say, for $n\geq\nd$). Indeed, suppose that in this setting, all the velocities $u_n$ ($n\geq 0$) are non-positive and the series of hits is infinite. Since $U_n>0$ with a probability larger than $1/2$ (provided that $\mu_n<+\infty$), there surely is a number $n\geq\nd$ such that $U_n>0$. Then the inequalities $u_{n-1}\leq 0$ and $M_n\geq m$ imply that
\[
u_n = \frac{(m-M_n)u_{n-1}+2M_nU_n}{m+M_n} > 0.
\]
The contradiction obtained shows that the probability of an infinite series of hits is $0$. Emphasize that the sequence $\{\mu_n\}_{n\geq 1}$ here is not assumed to be bounded.

These two situations seem to cover all practically meaningful choices of the ``collisional'' and ``Maxwellian'' masses.

If all the ``Maxwellian'' masses $\mu_1,\mu_2,\mu_3,\ldots$\ are finite then for any sequence of ``collisional'' masses $M_1,M_2,M_3,\ldots$, any mass $m$ of the atom, and any velocity $u_0<0$, one can choose a sequence of velocities $U_n>u_{n-1}$ for which the velocity $u_n$ will be negative for all $n$ and thus the corresponding series of hits will be infinite. Indeed, let $u_{n-1}<0$ for some $n\geq 1$. If we set, e.g.,
\[
U_n = \frac{3M_n-m}{4M_n}u_{n-1} > u_{n-1},
\]
we will obtain
\[
u_n = \frac{(m-M_n)u_{n-1}+2M_nU_n}{m+M_n} = \frac{u_{n-1}}{2} < 0.
\]
However, in the situations described in this section, the probability of such sequences of velocities $U_1,U_2,U_3,\ldots$\ is $0$ for any temperature $T_s>0$ of the surface.

\section{Infinite series of hits}
\label{infinite}

If the sequence $\{1/M_n\}_{n\geq 1}$ or $\{\mu_n\}_{n\geq 1}$ is unbounded then the probability of an infinite series of hits can be positive. In this section, we prove two statements illustrating this thesis.

1) For any mass $m>0$ of the atom, any sequence of ``collisional'' masses $\{M_n\}_{n\geq 1}$ with all terms less than $m$, any number $\eta<0$, any number $p$ in the range $0<p<1$, and any temperature $T_s>0$ there exists a sequence of finite ``Maxwellian'' masses $\{\mu_n\}_{n\geq 1}$ for which the series of hits of the atom against surface pseudoparticles is infinite with probability at least $p$ for any initial ``vertical'' velocity $u_0\leq\eta$.

Indeed, fix an arbitrary sequence of negative numbers $\kappa_0,\kappa_1,\kappa_2,\ldots$\ such that $\kappa_0\geq\eta$ and
\begin{equation}
\kappa_n > \frac{m-M_n}{m+M_n}\kappa_{n-1}
\label{kappa}
\end{equation}
for $n\geq 1$. Note that each such sequence strictly increases. Besides that, let $\delta_1,\delta_2,\delta_3,\ldots$\ be an arbitrary strictly decreasing sequence of numbers in the range $p<\delta_n<1$. Set $\chi_1=\delta_1<1$ and
\[
\chi_n = \delta_n/\delta_{n-1} < 1
\]
for $n\geq 2$. It is obvious that
\begin{equation}
\lim_{n\to\infty} \chi_n = 1.
\label{chi}
\end{equation}
Finally, for $n\geq 1$ define the positive numbers $\Gamma_n$ and $\mu_n$ by the relations
\begin{gather}
\Gamma_n = \frac{(M_n-m)\kappa_{n-1}+(m+M_n)\kappa_n}{2M_n},
\label{MnGamma} \\
\erf\left(\left[\frac{\mu_n}{2\kB T_s}\right]^{1/2}\Gamma_n\right) = \chi_n.
\label{Gammamu}
\end{gather}
The inequality $\Gamma_n>0$ follows from Eq.~\eqref{kappa}. Since $0<\chi_n<1$, the mass $\mu_n$ is also well defined.

Now let $u_{n-1}\leq\kappa_{n-1}$ and $U_n\leq\Gamma_n$ for some $n\geq 1$. Then $u_n\leq\kappa_n$. Indeed,
\[
u_n = \frac{(m-M_n)u_{n-1}+2M_nU_n}{m+M_n} \leq \frac{(m-M_n)\kappa_{n-1}+2M_n\Gamma_n}{m+M_n} = \kappa_n;
\]
in this calculation we used the inequality $m-M_n>0$ and the equality~\eqref{MnGamma}. Thus, if $u_0\leq\eta\leq\kappa_0$ and $U_n\leq\Gamma_n$ for all $n\geq 1$, then $u_n\leq\kappa_n<0$ for all $n\geq 0$, so that the series of hits will never terminate. On the other hand, the probability that $U_n\leq\Gamma_n$ for a certain $n$ is no less than the probability that $|U_n|\leq\Gamma_n$, and the latter probability is equal to $\chi_n$ in virtue of the formulas~\eqref{erf} and~\eqref{Gammamu}. Consequently, the probability that $U_n\leq\Gamma_n$ for all $n\geq 1$ is no less than
\[
\prod_{n=1}^\infty \chi_n = \delta_1 \prod_{n=2}^\infty \delta_n/\delta_{n-1} = \lim_{n\to\infty} \delta_n \geq p.
\]

The results of Sect.~\ref{finite} imply that if the sequence of ``collisional'' masses $\{M_n\}_{n\geq 1}$ under the hypotheses of the theorem just proven is bounded from below: $M_n\geq\Md$ for all $n\geq 1$ for a certain constant $\Md>0$, then the sequence of ``Maxwellian'' masses $\{\mu_n\}_{n\geq 1}$ is unbounded. This is clearly seen in the above calculations. If the sequence of ``collisional'' masses is bounded from below by a positive constant then the sequence $\{\Gamma_n\}_{n\geq 1}$ is bounded from above, so that it follows from Eqs.~\eqref{chi} and~\eqref{Gammamu} that $\lim_{n\to\infty} \mu_n = +\infty$.

2) For any mass $m>0$ of the atom, any sequence of finite ``Maxwellian'' masses $\{\mu_n\}_{n\geq 1}$, any number $\eta<0$, any number $p$ in the range $0<p<1$, and any temperature $T_s>0$ there exists a sequence of ``collisional'' masses $\{M_n\}_{n\geq 1}$ with all terms less than $m$ for which the series of hits of the atom against surface pseudoparticles is infinite with probability at least $p$ for any initial ``vertical'' velocity $u_0\leq\eta$.

Indeed, choose an arbitrary strictly increasing sequence of numbers $\kappa_0,\kappa_1,\kappa_2,\ldots$\ in the range $\eta\leq\kappa_n<0$ and an arbitrary strictly decreasing sequence of numbers $\delta_1,\delta_2,\delta_3,\ldots$\ in the range $p<\delta_n<1$. Define the sequence of numbers $\chi_1,\chi_2,\chi_3,\ldots$\ in the range $0<\chi_n<1$ in the same way as in the proof of statement~1). Finally, for $n\geq 1$ define the positive number $\Gamma_n$ by the relation~\eqref{Gammamu} and the mass $M_n$ by the relation~\eqref{MnGamma}:
\[
M_n = \frac{m(\kappa_n-\kappa_{n-1})}{2\Gamma_n-\kappa_{n-1}-\kappa_n}.
\]
Since $\kappa_{n-1}<\kappa_n<0$, one has $M_n>0$ and
\[
M_n = \frac{m(|\kappa_{n-1}|-|\kappa_n|)}{2\Gamma_n+|\kappa_{n-1}|+|\kappa_n|} < m.
\]
The further reasoning verbatim repeats the end of the proof of statement~1). Note that the equality~\eqref{MnGamma} and the inequality $\Gamma_n>0$ imply the inequality~\eqref{kappa}.

It follows from the results of Sect.~\ref{finite} that if the sequence of ``Maxwellian'' masses $\{\mu_n\}_{n\geq 1}$ under the hypotheses of the theorem just proven is bounded from above: $\mu_n\leq\mud$ for all $n\geq 1$ for a certain constant $\mud<+\infty$, then the sequence of ``collisional'' masses $\{M_n\}_{n\geq 1}$ is not bounded from below: it contains terms arbitrarily close to zero. The above calculations confirm this. If the sequence of ``Maxwellian'' masses is bounded from above then Eqs.~\eqref{chi} and~\eqref{Gammamu} imply that $\lim_{n\to\infty} \Gamma_n = +\infty$, whence $\lim_{n\to\infty} M_n = 0$.

By some misuse of language, one may say that an infinite series of hits in the composite encounter model corresponds to adsorption of the atom by the surface.

\section{Reciprocity issues}
\label{reciprocity}

In many problems of scattering theory, the so called \emph{reciprocity condition} plays a great role \cite{Cercignani72,Kuscer71,Liang18}. One of the forms of this condition is that if the velocities of the atoms incident on the surface have the three-dimensional Maxwellian distribution corresponding to the surface temperature $T_s$, then the velocities of the scattered atoms have the same Maxwellian distribution. However, in this paper we do not discuss a change in the ``horizontal'' component of the atom velocity upon an encounter with the surface. Hence, the reciprocity condition as applied to the model under consideration can be formulated as follows: if the ``vertical'' component $u_0<0$ of the velocity of the atoms incident on the surface has (up to the sign) the one-dimensional Maxwellian distribution corresponding to the surface temperature $T_s$ then the ``vertical'' component $\ubt>0$ of the velocity of the scattered atoms has the same Maxwellian distribution.

For some special values of the ``collisional'' and ``Maxwellian'' masses $M_n$ and $\mu_n$, the reciprocity condition is satisfied for the composite encounter model. For instance, let $M_1=\mu_1=+\infty$. Then $U_1=0$ and $\ubt=u_1=-u_0$ independently of the temperature $T_s$, of the mass $m$, and of the values of $M_n$ and $\mu_n$ for $n\geq 2$ (specular reflection), so that the distributions of the velocities $|u_0|$ and $\ubt$ coincide. However, generically the reciprocity condition is violated for the composite encounter model, even in the situations where a series of hits of the atom against surface pseudoparticles is finite with probability $1$, see Sect.~\ref{finite}. This signifies the presence of irreversible effects in scattering.

To be more precise, the following statement holds. For any temperature $T_s$ and mass $m$, any numbers $\Md>0$, $\eta<0$, $\kappa>0$, and any number $p$ in the range $0<p<1$ there exists a number $\mud>0$ such that if $\eta\leq u_0<0$, $M_1\geq\Md$, and $0<\mu_1\leq\mud$, then with probability at least $p$ the velocity $u_1$ of the atom after the first hit against a surface pseudoparticle satisfies the inequality $u_1\geq\kappa$ (and therefore $\ubt=u_1\geq\kappa$).

Indeed, the inequality
\[
u_1 = \frac{(m-M_1)u_0+2M_1U_1}{m+M_1} \geq \kappa
\]
is equivalent to the fact that
\begin{equation}
U_1 \geq \frac{(m+M_1)\kappa+(M_1-m)u_0}{2M_1} = \frac{\kappa+u_0}{2}+\frac{m(\kappa-u_0)}{2M_1}.
\label{uslovie}
\end{equation}
If $M_1\geq\Md$ and $\eta\leq u_0<0$ then the right-hand side of the inequality~\eqref{uslovie} is smaller than
\[
\Gamma = \max\left(\frac{\kappa}{2}+\frac{m(\kappa-\eta)}{2\Md}, \, -\eta\right) \geq -\eta \geq |u_0|.
\]
Define $\mud$ by the relation
\[
\erf\left(\left[\frac{\mud}{2\kB T_s}\right]^{1/2}\Gamma\right) = 1-p.
\]
Then for $\mu_1\leq\mud$ the probability that $|U_1| \geq \Gamma$ is no less than $p$. But since $\Gamma \geq |u_0|$, it follows from $|U_1| \geq \Gamma$ that $U_1\geq\Gamma$, whence the inequality~\eqref{uslovie} is also valid.

Now suppose that the velocities $|u_0|$ have the one-dimensional Maxwellian distribution corresponding to the surface temperature $T_s$. Choose an arbitrary number $\kappa>0$. Let $p$ be an arbitrary number in the range $0<p<1$. Take $\eta<0$ such that the inequality $u_0\geq\eta$ for the one-dimensional Maxwellian distribution with temperature $T_s$ holds with probability $p^{1/2}$:
\[
\erf\left(-\left[\frac{m}{2\kB T_s}\right]^{1/2}\eta\right) = p^{1/2}.
\]
After $\eta$ is determined, find $\mud>0$ such that the inequality $\ubt\geq\kappa$ for $u_0\geq\eta$ and $\mu_1\leq\mud$ is satisfied for the mass $M_1$ given with probability at least $p^{1/2}$. Then for $\mu_1\leq\mud$ the inequality $\ubt\geq\kappa$ holds with probability at least $p$ (provided that the velocities $|u_0|$ have the one-dimensional Maxwellian distribution corresponding to the temperature $T_s$ and the mass $M_1$ is fixed). But whenever the probability of the inequality $\ubt\geq\kappa$ occurring is too close to $1$, the distribution of velocities $\ubt$ definitely does not coincide with the one-dimensional Maxwellian distribution corresponding to the temperature $T_s$: for such a distribution, this probability is equal to
\[
1-\erf\left(\left[\frac{m}{2\kB T_s}\right]^{1/2}\kappa\right).
\]

\begin{figure*}
\includegraphics[width=0.75\textwidth]{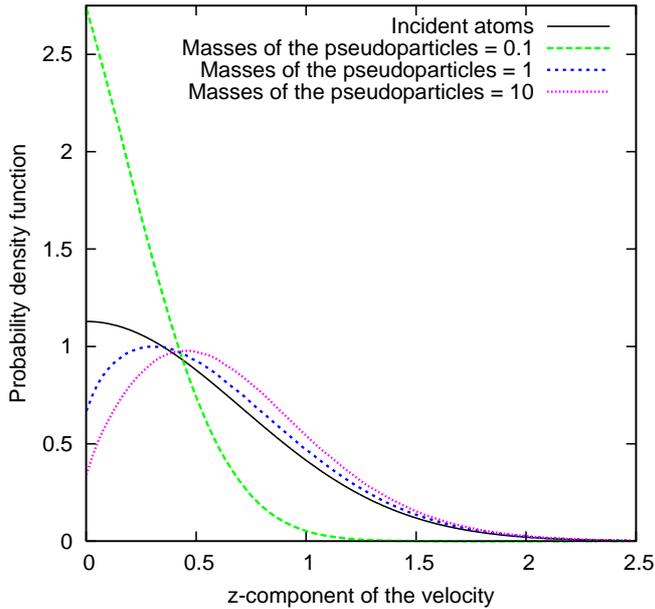}
\caption{Distributions of the ``vertical'' component $\ubt$ of the velocity of the scattered atoms at $m=1$, $\kB T_s=1/2$ for three values of the pseudoparticle masses $\cM$.}
\label{Fig1}
\end{figure*}

Computer simulations of the composite encounter model show that the reciprocity condition is violated even in the case where all the masses $M_n$ and $\mu_n$, $n\geq 1$, of the surface pseudoparticles are equal:
\[
M_1=\mu_1=M_2=\mu_2=M_3=\mu_3=\cdots = \cM<+\infty.
\]
As an example, for $m=1$ and three values of $\cM$ Fig.~\ref{Fig1} presents the distributions of the $z$-component $\ubt$ of the velocity of the scattered atoms under the condition that the $z$-component $u_0$ of the velocity of the atoms incident on the surface has (up to the sign) the one-dimensional Maxwellian distribution corresponding to surface temperature $T_s=1/(2\kB)$ (otherwise speaking, the probability density function of $u_0$ is $2\pi^{-1/2}e^{-u_0^2}$). For each value of $\cM$, we generated randomly $10^7$ encounters of the atom with the surface. It is seen in Fig.~\ref{Fig1} that for all the three values of $\cM$ (even for $\cM=1=m$), the distributions of $\ubt$ are very different from the Maxwellian distribution of $|u_0|$ (depicted in the figure by the solid black line). The mean number of hits of the atom against surface pseudoparticles within the framework of one encounter in our calculations was equal to 2.077, 1.149, and 1.033 for $\cM=0.1$, $1$, and $10$, respectively. The maximal number of hits we observed was equal to 12, 6, and 4, respectively. The results of Sect.~\ref{numberofhits} imply that, in fact, for any positive integer $\ell$ the number of hits is $\ell$ with a nonzero probability. As $\cM\to+\infty$, the distribution of the velocity $\ubt$ approaches the Maxwellian distribution of the velocity $|u_0|$.

It is shown in the recent paper \cite{Liang18} that the washboard model \cite{Tully90} does not satisfy the reciprocity condition either. The note \cite{Barwinkel93} discusses other scattering models (similar to the hard cube model) for which the reciprocity condition is violated.

A study of the distribution of the velocity $\ubt$ for given values of the velocity $u_0$, temperature $T_s$, and masses $m$, $M_n$, $\mu_n$ is the main task that arises when analyzing the composite encounter model (and which we hope to devote to subsequent publications).

\section{Conclusion}
\label{conclusion}

In conclusion, we explain how, using the composite encounter model, one can describe (within the framework of the quasiclassical trajectory method) scattering of a molecule consisting of several atoms (or ions in the case of ionic bonds) $\rA_1,\rA_2,\rA_3,\ldots$\ from a solid surface \cite{Azriel18,Azriel19}. To each atom $\rA_j$, one assigns its mass $m_j>0$, its radius $R_j>0$, a sequence of ``collisional'' masses $\{M_{n,j}\}_{n\geq 1}$ ($0<M_{n,j}\leq+\infty$), a sequence of ``Maxwellian'' masses $\{\mu_{n,j}\}_{n\geq 1}$ ($0<\mu_{n,j}\leq+\infty$), and a sequence $\{\cL_{n,j}\}_{n\geq 1}$ of laws of the change in the ``horizontal'' component of the atom velocity (of course, it is assumed that the choice of these masses and laws is physically motivated). The surface itself is the infinite ``horizontal'' plane $z=0$ and has some temperature $T_s>0$, while the atoms move in the half-space $z>0$.

The atom $\rA_j$ collides with the surface as a solid ball of radius $R_j$. As long as the distance from the nucleus of each atom to the surface (i.e., the $z$-coordinate of the nucleus of the atom) is greater than the radius of that atom, the motion of the system of atoms is simulated as the motion of classical point masses $m_1,m_2,m_3,\ldots$, their interaction with each other being described by a certain potential. As soon as the distance from the nucleus of some atom $\rA_j$ to the surface becomes equal to $R_j$, there occurs an \emph{instantaneous} encounter of this atom with the surface, which is described by the composite encounter model (under the assumption that the number of hits of the atom against surface pseudoparticles is finite). After this, the free motion of the system of atoms with the given potential resumes, until the next encounter of some atom with the surface or until the integration of the equations of motion is terminated.

According to this scheme, we have simulated dissociative scattering of molecules of potassium iodide ($\rA_1=\rK^+$, $\rA_2=\rI^-$) from a graphite surface \cite{Azriel18}. The masses $m_j$ of the ions corresponded to the atomic weights of potassium and iodine, the radii $R_j$ of the ions were set to be their crystal radii, whereas the interionic interaction potential was chosen on the basis of the data of the article \cite{Patil87}. It was assumed that the ``horizontal'' component of the velocity of each ion remains unaltered in encounters with the graphite surface (there are no tangential forces), the mass $\mu_{n,j}$ is equal to the carbon atom mass $\mC$ (to be more precise, it corresponds to the atomic weight of carbon) for all $n\geq 1$ and $j=1,2$, while the masses $M_{n,j}$ are equal to $nM_{1,j}$ (the graphite pseudoparticle taking part in the $n$th elastic hit of the ion against the surface within the framework of a given encounter, ``unites'' $n$ graphite layers parallel to the surface). Finally, we set $M_{1,j}=\Lambda_j\mC$ where $\Lambda_j$ is the mean number of carbon atoms in a graphite layer ``covered'' by a disc of radius $R_j$. The procedure for determining the coefficients $\Lambda_j$ is described in detail in the papers \cite{Azriel18,Azriel19}. According to Sects.~\ref{numberofhits} and~\ref{finite}, for such ``collisional'' and ``Maxwellian'' masses of graphite pseudoparticles, each encounter of the ions with the surface includes only a finite number of elastic hits (but any number of hits occurs with a nonzero probability).

We have carried out calculations for dissociative (and non-dissociative) scattering of KI molecules from a graphite surface \cite{Azriel18} for two values of the surface temperature, 298 and 523~K\@. The translational energy of molecules ranged between 4 and 14~eV, the angle between the axis of the molecular beam and the graphite surface ranged between $5^\circ$ and $85^\circ$, while the vibrational and rotational temperatures of molecules in the beam were set to be 600 and 70~K, respectively. The maximal number (which we observed) of elastic hits of the ion against graphite pseudoparticles in a single encounter of the ion with the surface was equal to 7, both for the potassium cation and the iodide anion.

The available experimental data on dissociation of KI molecules at the surface of pyrolytic graphite \cite{Azriel18} are rather limited. For instance, if $\bw_0$ denotes the velocity of the center of mass of the incident molecule and $\bw_j$ is the velocity of the ion formed ($j=1,2$), then in the experiment we only recorded the ions for which the vectors $\bw_0$ and $\bw_j$ are almost orthogonal and the plane they span is almost perpendicular to the graphite surface. Because of this, comparing the calculated dynamical characteristics of dissociation with the experimental results is not very informative. One will probably be able to reduce the discrepancies we observed \cite{Azriel18} by introducing tangential forces or by another choice of the masses $M_n$ and $\mu_n$.

\begin{acknowledgements}
The research was financed only from the state budget of the Russian Federation and was carried out within the framework of the Program of fundamental scientific research of the state academies of sciences of the Russian Federation for 2013--2020, the theme being ``Fundamental physical-chemical processes of the impact of energy objects on the environment and living systems''.
\end{acknowledgements}

\section*{Compliance with ethical standards}

{\small{\bfseries Conflict of interest} The authors declare that they have no conflict of interest.}

\end{document}